\documentclass[english,pra,showpacs,twocolumn]{revtex4-1}
\usepackage[T1]{fontenc}
\usepackage[latin9]{inputenc}
\usepackage{mathrsfs}
\usepackage{amsmath}
\usepackage{amssymb}
\usepackage{graphicx}

\makeatletter
\usepackage{txfonts}
\usepackage{dcolumn}% Align table columns on decimal point
\usepackage{bm}% bold math
\usepackage[colorlinks,
            linkcolor=blue,
            anchorcolor=blue,
            citecolor=red
            ]{hyperref}

\makeatother

\usepackage{babel}
\begin{document}

\title{Optimizing Quantum Teleportation and Dense Coding via Mixed Noise
Under Non-Markovian Approximation}

\author{Akbar Islam$^{1,2}$, An Min Wang$^{1,*}$}

\affiliation{$^{1}$Department of Modern Physics, University of Science and Technology
of China, Hefei 230026, China}
\email{anmwang@ustc.edu.cn}

\author{and Ahmad Abliz$^{2,\dagger}$}

\affiliation{$^{2}$School of Physics and Electronic Engineering, Xinjiang Normal
University, Urumqi, Xinjiang 830054, China}
\email{aahmad@126.com}

\begin{abstract}
Physicists are attracted to open-system dynamics, how quantum systems
evolve, and how they can protected from unnecessary environmental
noise, especially environmental memory effects are not negligible,
as with non-Markovian approximations. There are several methods to
solve master equation of non-Markovian cases, we obtain the solutions
of quantum-state-diffusion equation for a two qubit system using perturbation
method, which under influence of various types of environmental noises,
i.e., relaxation, dephasing and mix of them. We found that mixing
these two types of noises benefit the quantum teleportation and quantum
super-dense coding, that by introducing strong magnetic field on the
relaxation processes will enhance quantum correlation in some time-scale. 
\end{abstract}

\maketitle

\section{INTRODUCTION}

The global state of a composite system, if it cannot be written as
a product state of individual subsystems, implies that there is more
to the correlation between these subsystems than what first meets
the eye. Along with entanglement this has been used as a source of
various new discoveries, such as quantum cryptography \cite{Ekert1991},
quantum teleportation \cite{Bennett 1993}, and dense coding \cite{Bennet =000026 Wiesner 1992}.
However, the quantum states are fragile when encounter with the environmental
noises. For instance, environmental sensitivity (especially to noises)
is significant, and the reduction in efficiency of the quantum apparatus
cannot be ignored. Mitigating the degeneration, as well its effects,
has become one of the main focus of work today \cite{Neisen(2000)}. 

Since quantum information processing are usually disturbed by environmental
noises, various techniques have been developed for minimizing or eliminating
the degradation of entanglements. Most commonly used techniques are
quantum error correction \cite{Shor1995}\cite{Steane 1996b}, decoherence-free
subspace \cite{Lidar1998}\cite{Kwait2000}, and dynamical decoupling
\cite{Viola1999,YiXX1999,T.Yu(2004)}. 

Recent studies on two separable qubits coupled with single type of
noises \cite{Corn(2009),Jing(2013),Jing(2015),Beige(2000)}, have
shown that the quantum correlations between these qubits can be enhanced
or reduced through inducing other noises. This interesting phenomena
hints at an alternative approach to control dynamics of the open-quantum
systems using only noise itself. Jing et al. \cite{Jing(2018)} studied
a single qubit, simultaneously influenced by two types of noises,
using non-Markovian approximation, and succeeded in controlling relaxation
by dephasing noise. Their results showed that in the two- and three-level
atomic systems, non-Markovian relaxation processes can be repressed
using Markovian dephasing noise. In our study, we follow the steps
of Jing's work \cite{Jing(2018)}, increase the number of qubits from
one to two, and study how the quantum teleportation and super-dense
coding effected by theses noises. In this study, we rely on Quantum-State-Diffusion
(QSD) approach to solve master equations for the non-Markovian processes,
and carry on numerical analysis to see how different noise mixtures
affect quantum teleportation and super-dense coding. 

The rest of this paper is organized as follows. In Section II, we
introduce our model, a pair of qubits interacting with composite noises
in a common bath, and present the exact master equation's solution
for a two-qubit system simultaneously under the influence of two separate
noises by applying the QSD approach. In Section III, by analyzing
various parameters, we demonstrate how these different scenarios of
noises take their toll on super-dense coding and quantum teleportation.
And we also analyzed how the back-action induced by the strong system-bath
interaction effects mixed noise scenarios in non-Markovian approximation.
We present the conclusion to our work in Section \mbox{IV}. 

\section{MODEL AND METHOD}

\subsection{Model}

Let's construct an EPR pair model interacting with an environment
comprising a harmonic oscillator producing mixed noises in a common
bath. Based on the Jing's work \cite{Jing(2018)} for single qubit
model, the Hamiltonian for two-qubit system can be written as ($\hbar=1$)

\begin{equation}
H_{tot}=H_{sys}+H_{env}+H_{int}\label{eq:1}
\end{equation}
with

\begin{equation}
H_{sys}=\frac{1}{2}\left(\omega_{A}\sigma_{z}^{A}+\omega_{B}\sigma_{z}^{B}\right),\label{eq:2}
\end{equation}

\begin{equation}
H_{env}=\sum_{k}\omega_{k}b_{k}^{\dagger}b_{k},\label{eq:3}
\end{equation}
and
\begin{equation}
H_{int}\!\!=\!\frac{1}{2}\left[\xi\left(t\right)\sigma_{z}^{A}\!\!+\!\xi\left(t\right)\sigma_{z}^{B}\right]\!+\!\left(\sigma_{-}^{A}B_{A}^{\dagger}\!+\!h.c.\right)\!+\!\left(\sigma_{-}^{B}B_{B}^{\dagger}\!+\!h.c.\right).\label{eq:4}
\end{equation}
here, $\omega_{A}$ and $\omega_{B}$ are the frequencies of the qubit
$\mathit{A}$ and $\mathit{B}$, $\xi\left(t\right)$ is Gaussian
dephasing noise, satisfying $M\left[\xi\left(t\right)\right]=0$ ($M\left[\cdot\right]=\int\frac{d^{2}z}{\pi}e^{-\mid z\mid^{2}}$
denotes the ensemble average operation over all stochastic trajectories
noise $z_{t}^{*}$). The ensemble $M\left[\xi\left(t\right)\xi\left(t\right)\right]=\alpha\left(t-s\right)$
is a dephasing correlation function. It is clear that $\sigma_{z}$
and $\sigma_{\pm}=\left(\sigma_{x}\pm i\sigma_{y}\right)/2$ are Pauli
matrices, and $b_{k}$ and $\omega_{k}$ are respectively the annihilation
operators and eigenfrequencies of the $k$'th mode of the environment,
respectively. $B\equiv\sum_{k}g_{k}b_{k}$ serves as the collective
environmental operator describing the relaxation channel, while $g_{k}$
represents the coupling strength between the system and the environmental
modes. 

By performing a rotation frame on the Hamiltonian, with $e^{iS}$,
where $S=\sum_{k}\omega_{k}b_{k}^{\dagger}b_{k}^{\dagger}t+\frac{1}{2}\left[\Xi\left(t\right)\sigma_{z}^{A}+\Xi\left(t\right)\sigma_{z}^{B}\right]$,
and $\Xi\left(t\right)\equiv\int_{0}^{t}ds\xi\left(t\right)$, the
Hamiltonian can be rewritten as 

\begin{align}
H_{I} & =e^{iS}H_{tot}e^{-iS}-\dot{S}\nonumber \\
 & =H_{sys}+\sigma_{-}^{A}B^{\dagger}e^{-i\Xi\left(t\right)}+\sigma_{+}^{A}Be^{i\Xi\left(t\right)}+\sigma_{-}^{B}B^{\dagger}e^{-i\Xi\left(t\right)}+\sigma_{+}^{B}Be^{i\Xi\left(t\right)}\label{eq:5}
\end{align}
where $B\left(t\right)\equiv\sum_{k}g_{k}b_{k}e^{-i\omega_{k}t}$.
The correlation function of the non-Markovian relaxation environment
at the zero-temperature can be written as \cite{T.Yu(2004),L.Di=0000F3si(1998),T.Yu(1999),Chen(2014)}.
\begin{equation}
\beta\left(t-s\right)\equiv\langle B\left(t\right)B^{\dagger}\left(t\right)\rangle=\sum_{k}\mid g_{k}^{2}\mid e^{-i\omega_{k}\left(t-s\right)}\label{eq:6}
\end{equation}
If the initial state of the system $\mid\Psi_{0}\rangle$ and bath
(environment ) $\rho_{en}$ initially don't have any interaction,
then the initial density operator can be factorized as $\hat{\rho}_{tot}\left(0\right)=\mid\Psi_{0}\rangle\langle\Psi_{0}\mid\otimes\hat{\rho}_{en}$.
The stochastic Schrödinger equation for this model written as 
\begin{equation}
i\frac{\partial}{\partial t}\mid\Psi_{t}\rangle\equiv H_{I}\mid\Psi_{t}\rangle\label{eq:7}
\end{equation}
where $|\Psi_{t}\rangle$ is full state of the system environment
with a set of Bargmann coherent states $|z\rangle=|z_{1}\rangle\otimes|z_{2}\rangle\dots\otimes|z_{k}\rangle\otimes\dots$,
and $\psi_{t}\left(z^{*}\right)=\langle z\mid\Psi_{t}\rangle$ is
the stochastic wave function of the two-qubit system. The non-Markovian
stochastic Shrödinger equation, i.e. QSD equation \cite{L.Di=0000F3si(1998),T.Yu(1999)},
for two qubits interacting with mixed noises in a common bath is given
as follows: 

\begin{eqnarray}
\frac{\partial}{\partial t}\psi_{t}\left(z^{*}\right)\negthinspace & = & \!\left[-iH_{sys}\!+\!L_{A}z_{t}^{*}\!-\!L_{A}^{\dagger}\!\int_{0}^{t}ds\alpha\left(t,s\right)\frac{\delta}{\delta z_{s}^{*}}\!\right]\!\psi_{t}\left(z^{*}\right)\nonumber \\
 &  & +\left[\!L_{B}z_{t}^{*}\!-\!L_{B}^{\dagger}\!\int_{0}^{t}\!ds\alpha\left(t,s\right)\frac{\delta}{\delta z_{s}^{*}}\right]\!\psi_{t}\left(z^{*}\right)\label{eq:8}
\end{eqnarray}
where $L_{A}$ and $L_{B}$ are the system environment coupling operators.
According to Eq.(\ref{eq:5}) $L_{A}=\sigma_{-}^{A}$ and $L_{B}=\sigma_{-}^{B}$.
The composite noise can be presented as $z_{t}^{*}=-i\sum_{k}g_{k}^{*}z_{k}^{*}e^{i\omega_{k}t-i\Xi\left(t\right)}$,
which describes the combined effects of both relaxation (pure quantum
mechanical) and dephasing (semi-classical) noise processes on a two-level
two-qubit system in non-unitary evolution. These two types of noises
are assumed to be statistically independent \cite{Jing(2018)}. With
$\frac{\partial}{\partial z_{k}^{*}}=\int\frac{\partial z_{s}^{*}}{\partial z_{k}^{*}}\frac{\delta}{\delta z_{s}^{*}}ds=-i\int g_{k}e^{i\omega_{k}s}\frac{\delta}{\delta z_{s}^{*}}ds$
\cite{Chen(2014)}, the stochastic Shrödinger equation Eq.(\ref{eq:8})
can be transformed into a time-local form by replacing the functional
derivative in the integral with the time-dependent operator $O\left(t,s,z^{*}\right)$
(the $O$ operator) \cite{T.Yu(1999)}: 

\begin{equation}
\frac{\delta\psi_{t}\left(z^{*}\right)}{\delta z_{t}^{*}}=O\left(t,s,z^{*}\right)\psi_{t}\left(z^{*}\right).\label{eq:9}
\end{equation}
 In the Markov limit the $O$ operator equals to Lindblad operator.
By the ``consistency condition'' $\frac{\delta}{\delta z_{s}^{*}}\frac{\partial\psi_{t}\left(z_{t}^{*}\right)}{\partial t}=\frac{\partial}{\partial t}\frac{\delta\psi_{t}\left(z^{*}\right)}{\delta z_{s}^{*}}$\cite{L.Di=0000F3si(1998)},
the time evolution equation of the operator $O(t,s,z^{*})$ can be
written as: 
\begin{eqnarray}
\frac{\partial}{\partial t}O\left(t,s,z^{*}\right)\! & =\!\! & \left[-iH_{sys}\!+\!L_{A}z_{t}^{*}\!-\!L_{A}^{\dagger}\bar{O}\left(t,z^{*}\right)\!,O\left(t,s,z^{*}\right)\right]\nonumber \\
\!\!\negthinspace & \!\!\negthinspace & \!\!\negthinspace+\!\!\left[\!L_{B}z_{t}^{*}\!\!-\!\!L_{B}^{\dagger}\bar{O}\!\left(t,\!\!z^{*}\right)\!,\!O\!\left(t\!,\!s\!,\!z^{*}\right)\!\right]\!\!-\!\!L^{\dagger}\!\frac{\delta}{\delta z_{s}^{*}}\bar{O}\!\left(t\!,\!z^{*}\right)\label{eq:10}
\end{eqnarray}

where $L=L_{A}+L_{B}$, $\bar{O}\left(t,z^{*}\right)\equiv\int_{0}^{t}dsG\left(t-s\right)O\left(t,s,z^{*}\right)$
, and from the two-qubit interaction model, the $O\left(t,s,z^{*}\right)$
can be written as follows: 

\begin{equation}
O\left(t,s,z^{*}\right)=O_{0}\left(t,s\right)+i\int_{0}^{t}ds_{1}z_{s_{1}}^{*}O_{1}\left(t,s,s_{1}\right)+\dots,\label{eq:11}
\end{equation}
$O_{0}\left(t,s\right)$ and $O_{1}\left(t,s,s_{1}\right)$ correspond
to the zeroth- and first-order noise components (higher orders can
be ignored for two-qubit cases) \cite{L.Di=0000F3si(1998),T.Yu(1999)}.
By using the consistency condition $O\left(t,s,z^{*}\right)$ can
be expanded as: 

\begin{eqnarray*}
\!\!\negthinspace O_{0}\left(t,s\right)\! & = & \!f_{1}\left(t,s\right)\sigma_{-}^{A}\!+\!f_{2}\left(t,s\right)\sigma_{-}^{B}\!+\!f_{3}\left(t,s\right)\sigma_{z}^{A}\sigma_{-}^{B}\!+\!f_{4}\left(t,s\right)\sigma_{-}^{A}\sigma_{z}^{B}
\end{eqnarray*}
\begin{equation}
O_{1}\left(t,s,s_{1}\right)=f_{5}\left(t,s,s_{1}\right)\left(2\sigma_{-}^{A}\sigma_{-}^{B}\right)\label{eq:12}
\end{equation}
Where values of $f_{1\sim5}$ are time-dependent, and are noise free.
Substituting the $O$ operator with its expanded form in Eq.(\ref{eq:10}),
we can obtain the partial differential equations that determine the
coefficients of the $O$ operator: 

\begin{align}
\!\!\frac{\partial}{\partial t}f_{1}\left(t,s\right) & \!=\!\!\left(i\omega_{A}\!\!+\!F_{1}\!\!+\!F_{3}\right)f_{1}\!+\!\left(\!F_{4}\!-\!F_{1}\right)f_{3}\!+\!\left(F_{3}\!+\!F_{4}\right)f_{4}\!-\!iF_{5},\nonumber \\
\!\!\frac{\partial}{\partial t}f_{2}\left(t,s\right) & \!=\!\!\left(i\omega_{B}\!+\!F_{2}\!+\!F_{4}\right)f_{2}\!+\!\!\left(F_{4}\!+\!F_{3}\right)f_{3}\!+\!\left(F_{3}\!-\!F_{2}\right)f_{4}\!-\!iF_{5},\nonumber \\
\frac{\partial}{\partial t}f_{3}\left(t,s\right) & \!=\!\!\left(i\omega_{B}\!+\!F_{2}\!+\!F_{4}\right)f_{3}\!+\!\!\left(F_{3}\!+\!F_{4}\right)f_{2}\!+\!\left(F_{3}\!-\!F_{2}\right)f_{1}\!-\!iF_{5},\nonumber \\
\frac{\partial}{\partial t}f_{4}\left(t,s\right) & \!=\!\!\left(i\omega_{A}\!+\!F_{1}\!+\!F_{3}\right)f_{4}\!+\!\left(\!F_{3}\!+\!F_{4}f_{1}\right)\!+\!\left(F_{4}\!-\!F1\right)f_{2}\!-\!iF_{5},\nonumber \\
\frac{\partial}{\partial t}f_{5}\left(t,s,s_{1}\right) & \!\!=\!\!\left(2i\omega_{A}\!\!+\!\!2i\omega_{B}\!+\!\!F_{1}\!+\!\!F_{2}\!+\!\!F_{3}\!+\!\!F_{4}\right)f_{5}\!\!+\!\!F_{5}\left(f_{1}\!+\!\!f_{2}\!-\!\!f_{3}\!-\!\!f_{4}\right),\label{eq:13}
\end{align}
where $F_{j}\left(t\right)=\int_{0}^{t}dsG\left(t,s\right)f_{j}\left(t,s\right)$
$ $$\left(j=1,2,3,4\right)$, and $F_{5}\left(t,s_{1}\right)=\int_{0}^{t}dsG\left(t-s\right)f_{5}\left(t,s,s_{1}\right)$,
with the initial conditions for equation above:

\begin{eqnarray}
f_{1}\left(t,s=t\right) & = & 1\nonumber \\
f_{2}\left(t,s=t\right) & = & 1\nonumber \\
f_{3}\left(t,s=t\right) & = & 0\nonumber \\
f_{4}\left(t,s=t\right) & = & 0\nonumber \\
f_{5}\left(t,s=t,s_{1}\right) & = & 0\nonumber \\
f_{5}\left(t,s,s_{1}=t\right) & = & -i\left[f_{3}\left(t,s\right)+f_{4}\left(t,s\right)\right].\label{eq:14}
\end{eqnarray}

After obtain these coefficients $f_{1\sim5}$ and $F_{1\sim5}$, determined
by Eq.(\ref{eq:13}), provide the answers to Eq.(\ref{eq:12}). The
density operator of the system at final state is defined \cite{T.Yu(2004),L.Di=0000F3si(1998),T.Yu(1999)}
as $\rho_{t}=\mathscr{M}\left[P_{t}\right]=\mathscr{M}\left[|\psi_{t}\left(z^{*}\right)\rangle\langle\psi_{t}\left(z^{*}\right)|\right]$,
by applying Novikov-type theorem $\mathscr{M}\left[z_{t}P_{t}\right]=\int_{0}^{t}dsG\left(t-s\right)\mathscr{M}\left[O\left(t,s,z_{t}^{*}\right)P_{t}\right]$
\cite{L.Di=0000F3si(1998)}. We can numerically solve the evolution
of Shrödinger equation Eq.(\ref{eq:8}), leading to the solution of
the master equation below \cite{L.Di=0000F3si(1998)}
\begin{equation}
\partial_{t}\rho_{t}=-i\left[H_{sys},\rho_{t}\right]+\left[L,\mathscr{M}\left[P_{t}\bar{O}^{\dagger}\right]\right]-\left[L^{\dagger},\mathscr{M}\left[\bar{O}P_{t}\right]\right]\label{eq:15}
\end{equation}

For environmental noise $z_{t}^{*}$ the correlation function can
be written as \cite{Jing(2018)}
\begin{eqnarray}
G\left(t-s\right) & = & \mathscr{M}\left[z_{t}^{*}z_{s}^{*}\right]\nonumber \\
 & = & \sum|g_{k}^{2}|e^{i\left[\Xi\left(t\right)-\Xi\left(s\right)\right]-i\omega_{k}\left(t-s\right)}\nonumber \\
 & = & \beta\left(t-s\right)\left[-\int_{0}^{t}dt_{1}\int_{0}^{t}dt_{2}\alpha\left(t_{1}-t_{2}\right)\right]\label{eq:16}
\end{eqnarray}
when only relaxation noise is present, $G\left(t-s\right)$ is reduced
to $\beta\left(t-s\right)$. For simplicity, both the dephasing and
relaxation noises are chosen as Ornstein-Uhlenbeck (OU) noises \cite{Jing(2018)}
depicted by the correlation functions $\alpha\left(t-s\right)=\frac{\Gamma_{\alpha}\gamma_{\alpha}}{2}e^{-\gamma_{\alpha}\mid t-s\mid}$
and $\beta\left(t-s\right)=\frac{\Gamma_{\beta}\gamma_{\beta}}{2}e^{-\gamma_{\beta}\mid t-s\mid}$,
where $\gamma_{\alpha}$ and $\gamma_{\beta}$ are inverses of the
memory capacities of the relevant noise or environment, and $\Gamma_{\alpha}$
and $\Gamma_{\beta}$ are their coupling strengths. The combined noise
correlation function can be expressed as
\begin{alignat}{1}
G\left(t-s\right)=\frac{\Gamma_{\beta}\gamma_{\beta}}{2}e^{-\gamma_{\beta}\mid t-s\mid}\exp\left\{ -\frac{\Gamma_{\alpha}}{2}\left[\left(t-s\right)+\frac{e^{-\gamma_{\alpha}\left(t-s\right)}-1}{\gamma_{\alpha}}\right]\right\} \label{eq:17}
\end{alignat}

It is obvious that the composite correlation function $G\left(t-s\right)$
is not in a linear exponential form, as Jing mentioned in \cite{Jing(2018)}.
Furthermore, the combining two OU noises will not yield another OU
noise, except there is a Markovian limit for either of these noises,
for example; when $\gamma{}_{\alpha}\to\infty$, meaning the dephasing
noise is Markovian, $\alpha\left(t-s\right)=\Gamma_{\alpha}\delta\left(t-s\right)$.
By using above conditions Eq.(\ref{eq:17}) reduces to

\begin{equation}
G\left(t-s\right)=\frac{\widetilde{\Gamma}_{\beta}\tilde{\gamma_{\beta}}}{2}\exp\left[-\tilde{\gamma_{\beta}}\mid t-s\mid\right]\label{eq:18}
\end{equation}
where $\widetilde{\Gamma}_{\beta}=r\Gamma_{\beta}$, $r=\gamma_{\beta}/\tilde{\gamma}_{\beta}$
and $\tilde{\gamma}_{\beta}=\gamma_{\beta}+\Gamma_{\alpha}/2$. With
$\tilde{\gamma}_{\beta}>\gamma_{\beta}$, the memory effect becomes
weaker, and since $r<1$, and $\widetilde{\Gamma}_{\beta}<\Gamma_{\beta}$
the coupling strength reduces. The parameters of non-Markovian memory
capacities $\gamma_{\alpha}$ and $\gamma_{\beta}$ for relaxation
and dephasing noises are from separable distinctive sources. This
leads to the composite modified noise correlation function becoming
shorter when compared to the purely non-Markovian relaxation noise
correlation function $\beta\left(t-s\right)$. The exponential component
of the correlation function will change the probability of restoring
the qubit to its original status through backflow from the environment
during the non-Markovian process. 

\subsection{Capacity of quantum super-dense coding}

Super-dense coding was proposed by Bennet and Wiesner in 1992 \cite{Bennet =000026 Wiesner 1992}.
He suggested that, when a sender performs a local unitary transformation
$U_{i}\in U\left(d\right)$ on the qubit in hand with $\rho$, $d$
is the dimension of the quantum system, the shared quantum systems
of both sender and receiver will be put into a state $\rho_{i}=\left(U_{i}\otimes I_{d}\right)\rho\left(U_{i}\otimes I_{d}\right)$,
with the probabilities of $p_{i}$ ($i=0,1,\dots,i_{max}$). For a
two-level quantum state $\mid j\rangle=\mid0\rangle,\mid1\rangle$,
the local unitary transformations $U_{i}$ are chosen as

\begin{align}
U_{00}\mid j\rangle & =\mid j\rangle\nonumber \\
U_{01}\mid j\rangle & =\mid j+1\rangle\left(mod2\right)\nonumber \\
U_{10}\mid j\rangle & =e^{\sqrt{-1}\frac{2\pi}{2}j}\mid j\rangle\nonumber \\
U_{11}\mid j\rangle & =e^{\sqrt{-1}\frac{2\pi}{2}j}\mid j\rangle\label{eq:19}
\end{align}

Super-dense coding has made it possible to transmit more information,
by combining quantum entanglement states with quantum channels, than
classical communication allowed previously. The information volume
capability of super-dense coding is scaled by the Holevo quantity
\cite{Holevo1973} . This is recognized as capacity

\begin{equation}
\chi=S\left(\overline{\rho}\right)-S\left(\rho\right)\label{eq:20}
\end{equation}
where $S\left(\rho\right)=-Tr\left(\rho\log_{2}\rho\right)$ is the
von-Neumann entropy and $\overline{\rho}=\sum_{i=0}^{i_{max}}p_{i}\rho_{i}$
is the average density matrix of the unified ensemble. 

\subsection{Fidelity of quantum teleportation}

Since early 1993, when Bennett et al. proposed the idea of quantum
teleportation \cite{Bennett 1993}, many experiments have successfully
realized quantum teleportation \cite{Boschi1998,Marcikic2003,Xiao-Song Ma2012,Bouwmeester1997}.
Fidelity is a popular method for measuring the distance between two
quantum states, to find out how the environment memory effects EPR
pair shared by Alice and Bob\cite{Bennet =000026 Wiesner 1992,Jozsa1994}.
For the information to be transmitted, the qubit can be expressed
as a vector on the Bloch sphere\cite{Neisen(2000)}

\begin{equation}
\mid\psi\rangle_{in}=\cos\frac{\theta}{2}\mid0\rangle+e^{i\phi}\sin\frac{\theta}{2}\mid1\rangle\label{eq:21}
\end{equation}
where $0\le\theta\le\pi$, $0\le\phi\le2\pi$ are the polar and azimuthal
angles, respectively. Then the initial state can be expressed as $\rho_{in}=\mid\psi\rangle_{in}\langle\psi\mid$.
After quantum teleportation, the output state will be\cite{Lombardi2002,Bowen(2001)}

\begin{equation}
\rho_{out}^{m}=\sum_{k=0}^{3}\langle\Psi_{Bell}^{k\oplus m}\mid\rho_{t}\mid\Psi_{Bell}^{k\oplus m}\rangle\otimes\sigma^{k}\rho_{in}\sigma^{\dagger k}\label{eq:22}
\end{equation}
where $m\left(m=0,1,2,3\right)$ indicates the Bell states of quantum
channels (EPR pair), which are $\mid\varphi_{Bell}^{0,3}\rangle=\frac{1}{2}\left(\mid00\rangle\pm\mid11\rangle\right)$
and $\mid\varphi_{Bell}^{1,2}\rangle=\frac{1}{2}\left(\mid01\rangle\pm\mid10\rangle\right)$.
Thus we can obtain $\rho_{t}$ from Eq. (\ref{eq:15}). By measuring
the distance between the input states $\rho_{in}$ and output states
$\rho_{out}$, the fidelity can be obtained as $\mathscr{\mathcal{F}^{m}=\langle\psi_{in}\mid\rho_{out}^{m}\mid\psi_{in}\rangle}$.
It's obvious that fidelity is dependent on random angles such as $\theta$
and $\phi$ of the input state. Generally, the exact state to be sent
via teleportation is unknown, thus calculating the average fidelity
is reasonable. The average fidelity $\overline{\mathcal{F}}$ of teleportation
can be written as\cite{Zhang Guo Feng(2007)} .

\begin{equation}
\overline{\mathcal{F}}=\frac{1}{4\pi}\int_{0}^{2\pi}d\phi\int_{0}^{\pi}\mathcal{F}sin\theta d\theta\label{eq:23}
\end{equation}

\section{Analyze of noise control effeminacy }

The quantum teleportation and super-dense coding plays key roll in
quantum teleportation and quantum communication. It will be ideal
to enhance them without causing complication. For that goal, we blend
the non-Makovian relaxation noise with dephasing noise, hope this
can provides some useful thoughts. We mainly focus on the EPR pairs,
under the influence of various noise types; non-Markovian pure relaxation
noise, Markovian pure dephasing noise and non-Markovian relaxation
noise mixed with the Markovian dephasing noise \cite{Jing(2018)}.
By using our exact non-Markovian master equation Eq.(\ref{eq:15}),
we obtain the capacity $\chi$ of super-dense coding for initial state
$\mid\psi\rangle=1/\sqrt{2}\left(\mid01\rangle+\mid10\rangle\right)$
, and average fidelity $\overline{\mathcal{F}}$ of teleportation
which the channel is $\mid\psi\rangle$ too. 

\begin{figure}
\includegraphics[width=8cm]{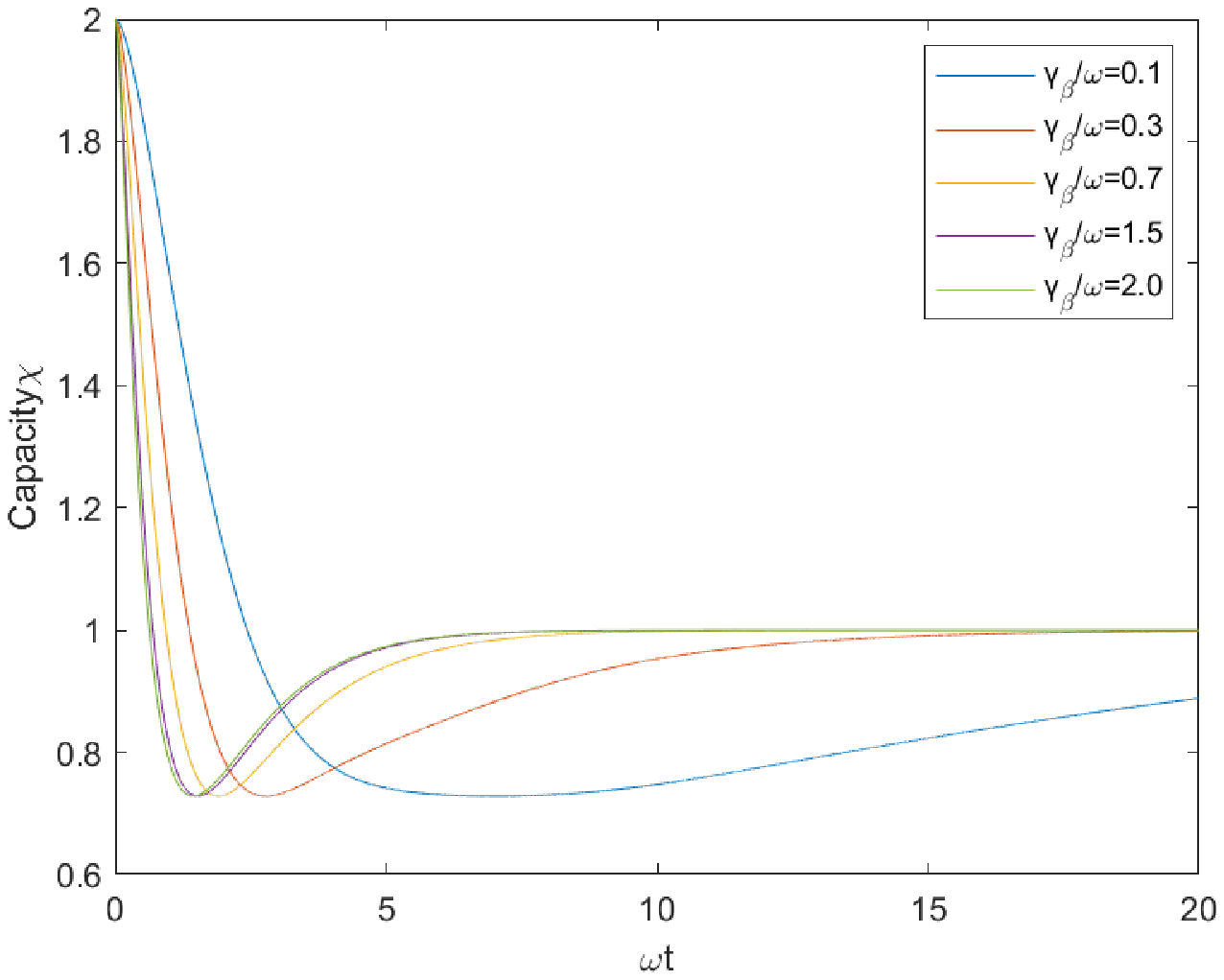}\label{fig1a}fig.1a

\includegraphics[width=8cm]{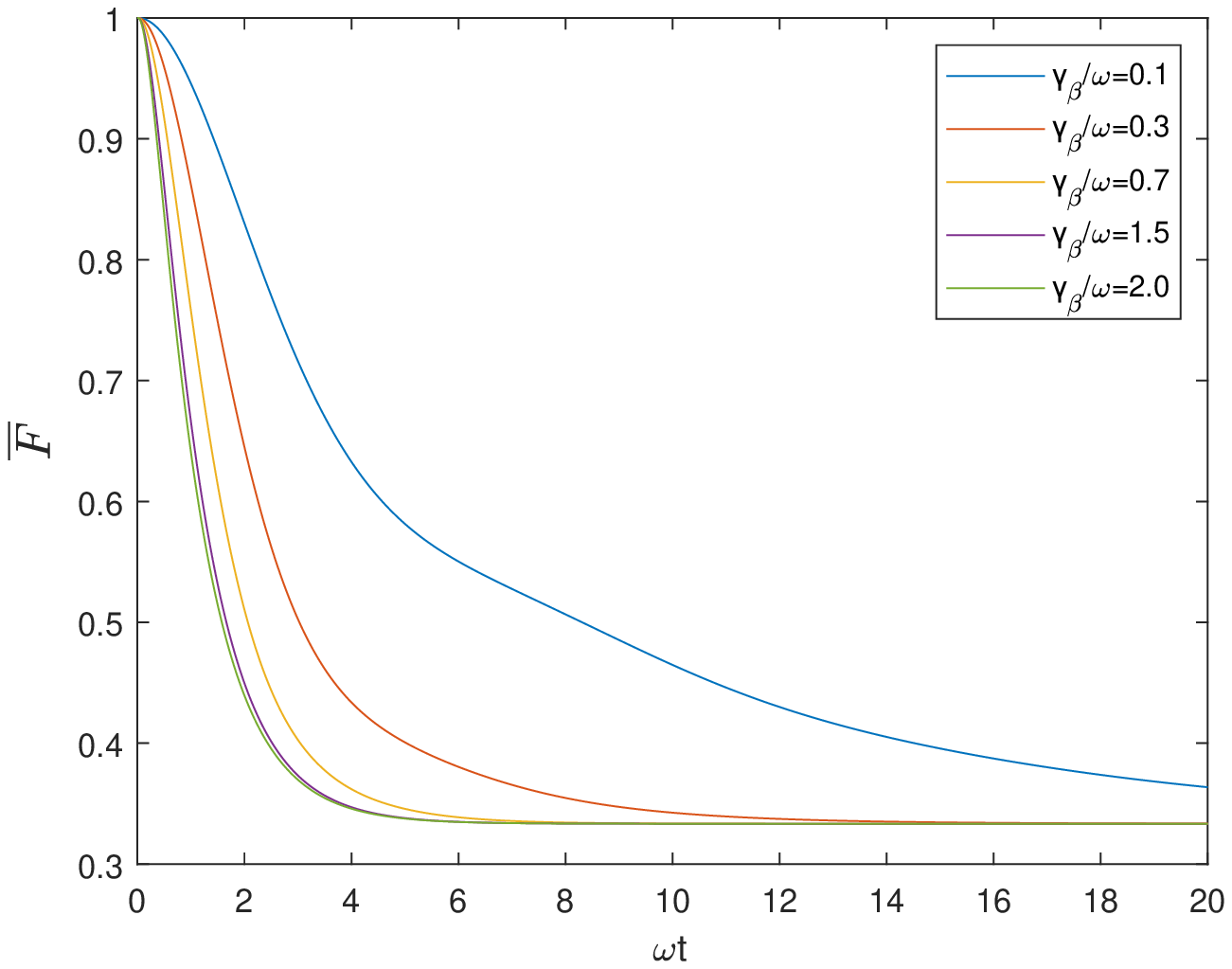}\label{fig1b}fig.1b

\caption{(Color online) The capacities $\chi$ of super-dense coding and the
average fidelity $\overline{\mathcal{F}}$ of the teleportation of
our system, under the influence of composite non-Markovian noise processes,
when the dephasing noise coupling strength is fixed as $\Gamma_{\alpha}/\omega=1$,
with various memory capacities $\gamma_{\beta}/\omega=0.1$,$0.3$,$0.7$,$1.5$,$2.0$,
which in blue, red, yellow, purple and green colors, respectively.
\label{fig1}}
\end{figure}
 To understand the degree of entanglement generation in non-Markovian
relaxation process encounter with Markovian dephasing noise, we plot
the capacities $\chi$ and average fidelity $\overline{\mathcal{F}}$
of our quantum system with the increase of inverse memory capacity
$\gamma_{\beta}/\omega$ (from $0.1\sim2.0$), in Fig.(\ref{fig1}a)-(\ref{fig1}b).
When there is long system-environment memory ($\gamma_{\beta}/\omega=0.1$),
the backflow of information into quantum systems delays the decoherence
between the qubits, improves the $\chi$ and $\overline{\mathcal{F}}$.
With the inverse memory time parameter $\gamma_{\beta}$ increases,
decoherence gets faster, and less the $\chi$ and $\overline{\mathcal{F}}$.
The higher capacities and fidelity rely on longer memory time ($1/\gamma_{\beta}$)
shows robustness of non-Markovian composite noises on entanglement
generation compare to the near Markovian or classical processes. Again,
this proves the suggestion of Yu Ting et al. \cite{T.Yu(2004),Jing(2013),Jing(2015),Jing(2018),T.Yu(1999),YuTing2004,LuoS.L.(2010)},
that with an increase in non-Markovian noise memory capacity, quantum
systems become more tolerant toward decoherence induced by the environment;
information from the environment flows back into the system itself
and restores some of the coherence between the qubits. 

\begin{figure}
\includegraphics[width=8cm]{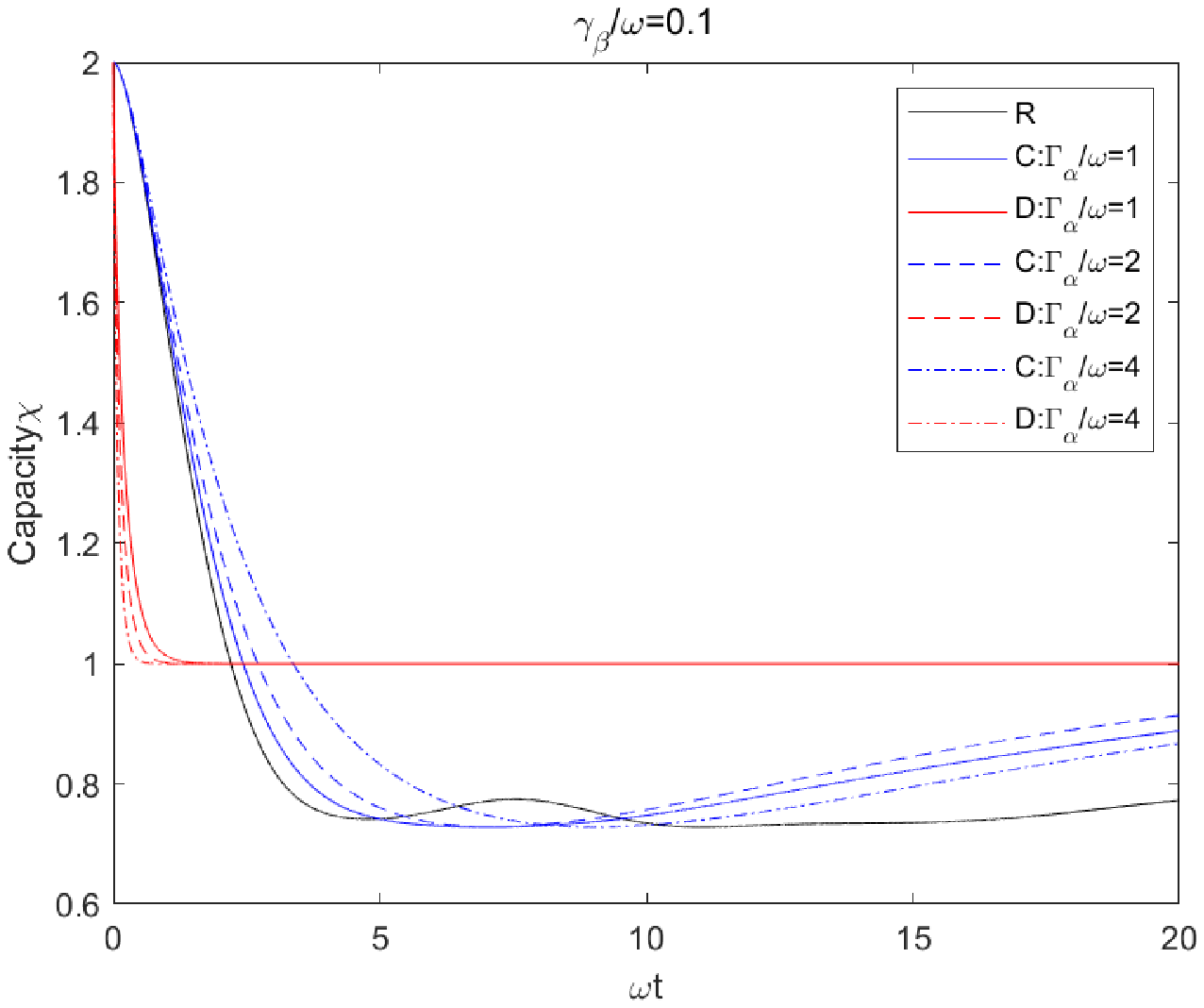}\label{fig2a}fig.2a

\includegraphics[width=8cm]{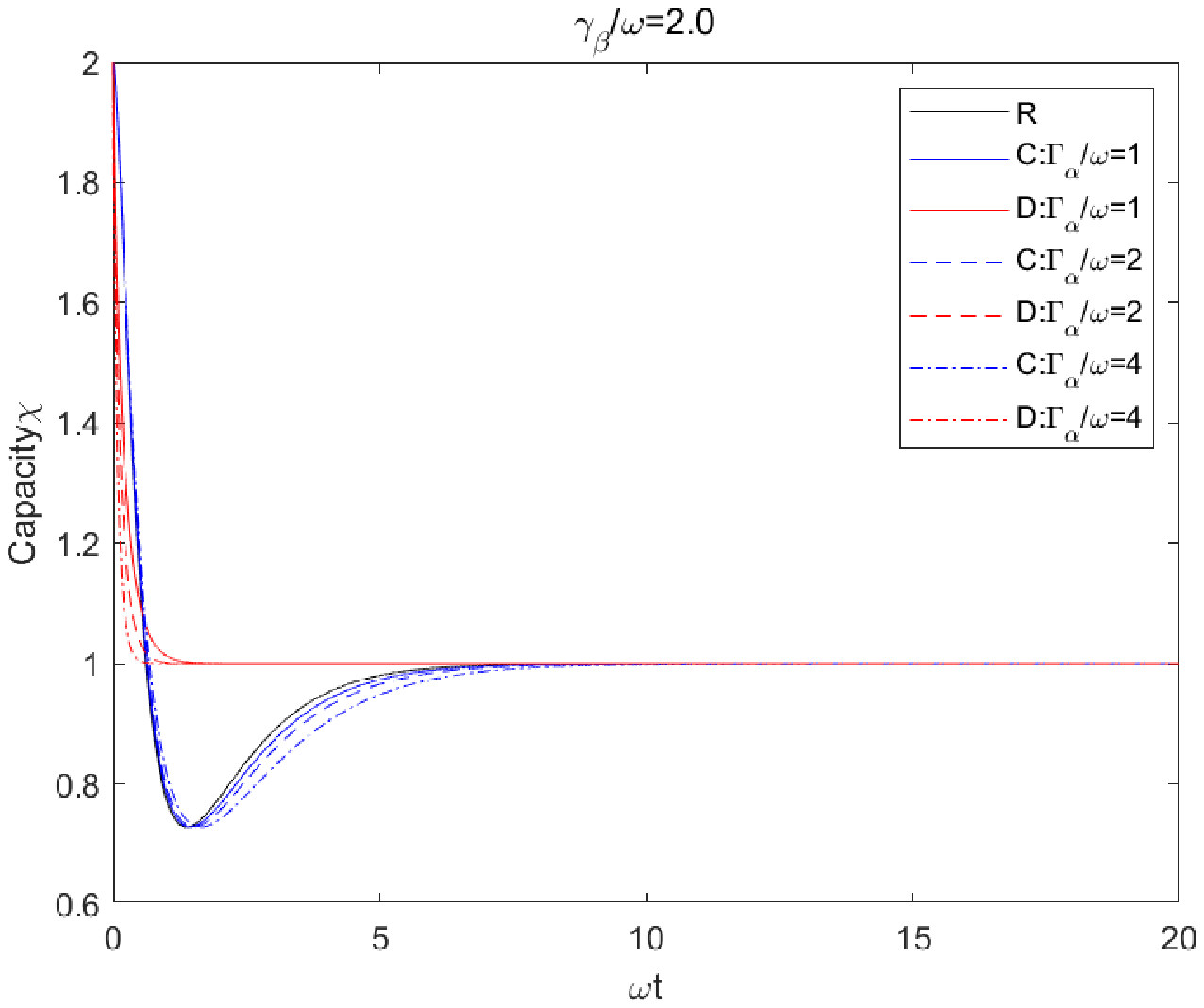}\label{fig2b}fig.2b

\caption{(Color online) Capacities $\chi$ of the two-level two-qubit quantum
systems influenced by Markovian dephasing noise in a non-Markovian
relaxation process, with various memory capacities; (a)$\gamma_{\beta}/\omega=0.1,$
(b)$\gamma_{\beta}/\omega=2.0$. The R, D, and C are the capacities
$\chi$ under non-Markovian pure relaxation, Markovian pure dephasing,
and composite non-Markovian noises, respectively. $\Gamma_{\beta}/\omega=1,2,4$
are fixed parameters presenting dephasing noise coupling strengths,
drawn with solid, dashed, and dot-dashed lines. \label{fig2}}
\end{figure}
The next step is to learn how Markovian dephasing noise contributes
to information preservation, which introduced to quantum super-dense
coding processes by a magnetic field, when non-Markovian relaxation
occurs. We compare the differences of capacities $\chi$ of super-dense
coding with the various noise types; the solid black curve labeled
with R shows the pure relaxation processes, the red curves labeled
with D show the Markovian dephasing noise processes, the blue curves
labeled with C show the composite noise processes. The Fig.\ref{fig2}(a,b)
are organized as increase of $\gamma_{\beta}$, and in each figures
the dephasing decoupling rates $\Gamma_{\alpha}/\omega$ are increased
from 1\textasciitilde 4. Fig.\ref{fig2}a shows, when the systems
under go strong non-Markovian processes with inverse memory capacity
$\gamma_{\beta}/\omega=0.1$. All three groups of different colored
curves with the same dephasing coupling strengths $\Gamma_{\alpha}$
can be divided in to two regions. In a moderate-time scale $0<\omega t<\tau$,
with the value $2.5<\tau<3.5$, the systems under the mixture noises
C have the higher capacities than the systems under pure dephasing
D or relaxation R noises. The increasing $\Gamma_{\alpha}$ yield
even higher capacities and longer delay of $\tau$. For the rest of
the timescale $3.5<\omega t$, the capacities $\chi$ of system under
mixture noises C is higher than that system under the pure relaxation
noises most of the time, but lower than the pure dephasing processes
D. When the system embedded in mixture noises environment, which are
blue curves C, the capacities $\chi$ of the system enhanced comparatively
with the increase of $\Gamma_{\alpha}$. Hence our noise control protocol
for two-qubit systems works for canceling relaxation effects in dissipation
processes. 

When the relaxation noise is a near Markovian with $\gamma_{\beta}/\omega=2.0$
\cite{T.Yu(2004),Jing(2018),ZhaoXY(2011)}, we obtain Fig.\ref{fig2}b.
Due to a smaller non-Markovian memory capacity $1/\gamma_{\beta}$,
all three noise type scenarios show significantly faster decay compared
to Fig.\ref{fig2}a. Its hard to distinguish the pure relaxation R
from mixture noises C at smaller dephasing decoupling rate $\Gamma_{\alpha}/\omega=1$,
but the higher dephasing decoupling rates $\Gamma_{\alpha}/\omega=2,4$
benefit the mixture noise. By introducing a dephasing noise in the
near Markovian relaxation noise, it improves capacities of mixture
noises. However, under the high dephasing decoupling rate $\Gamma_{\alpha}/\omega=4$,
the capacities $\chi$ of mixture noises are slightly higher than
the system under the pure dephasing noise D in a very short time scale
$0<\omega t<1.0$. 

The Fig.\ref{fig2}(a,b) show that, by introducing a dephasing noise
into two-qubit non-Markovian relaxation processes, the decay of super-dense
coding capacities $\chi$ can be delayed. The stronger dephasing noise
coupling strength $\Gamma_{\alpha}$, i.e. strong magnetic field,
delays the decay even noticeably. Adding the differences made by non-Markovian
memory capacity $1/\gamma_{\beta}$ into consideration; with the help
of high memory capacity $1/\gamma_{\beta}$ and strong dephasing noise
coupling strength $\Gamma_{\alpha}$ significantly delay the decay
of capacities $\chi$ of super-dense coding (in Fig\ref{fig2}a blue
dot-dash lines), than when they work separately. Thus, by inducing
a strong dephasing noise into the non-Markovian relaxation processes,
delay the decoherence and help the system to performs better on super-dense
coding. 

\begin{figure}
\includegraphics[width=8cm]{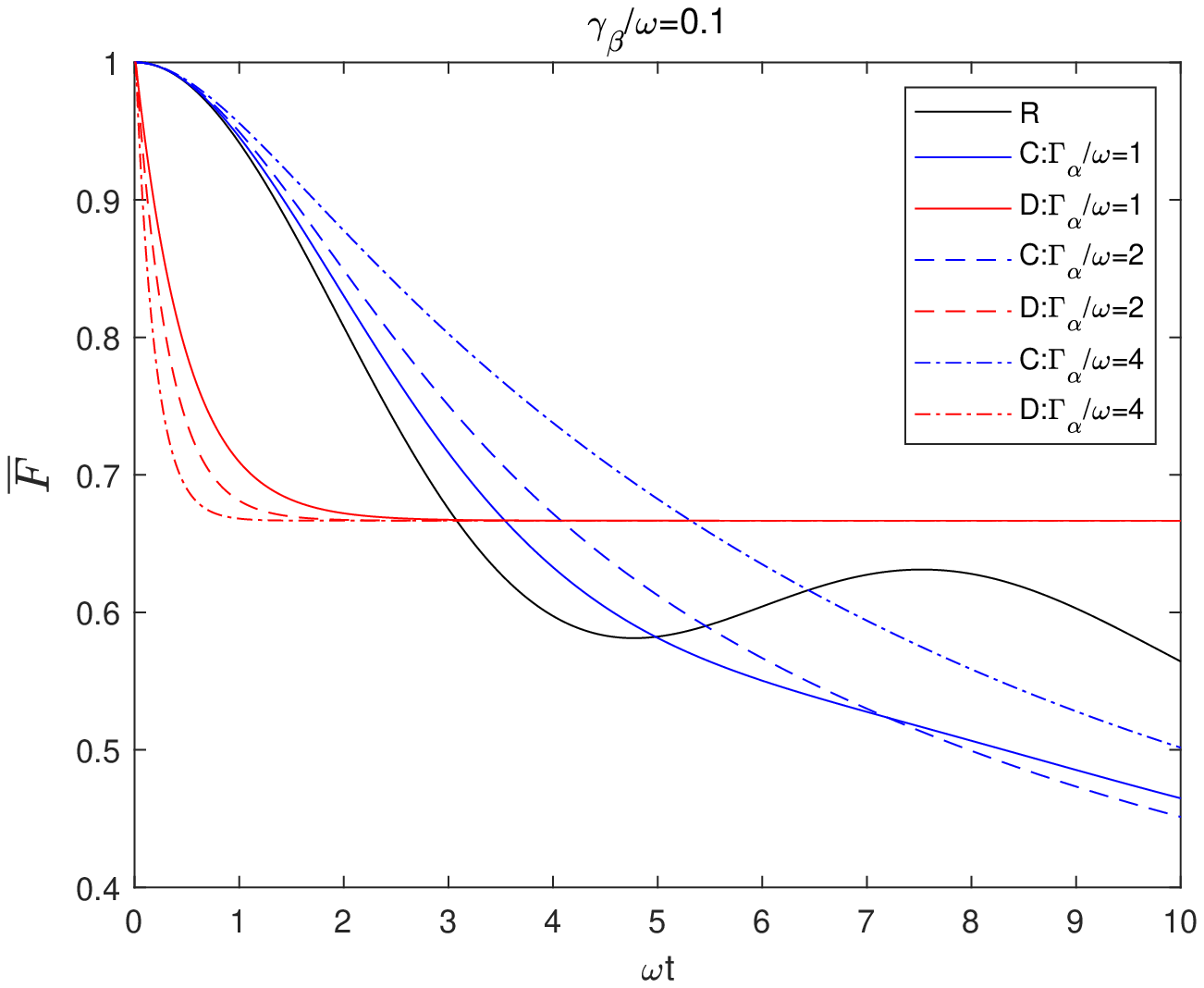}\label{fig3a}fig.3a

\includegraphics[width=8cm]{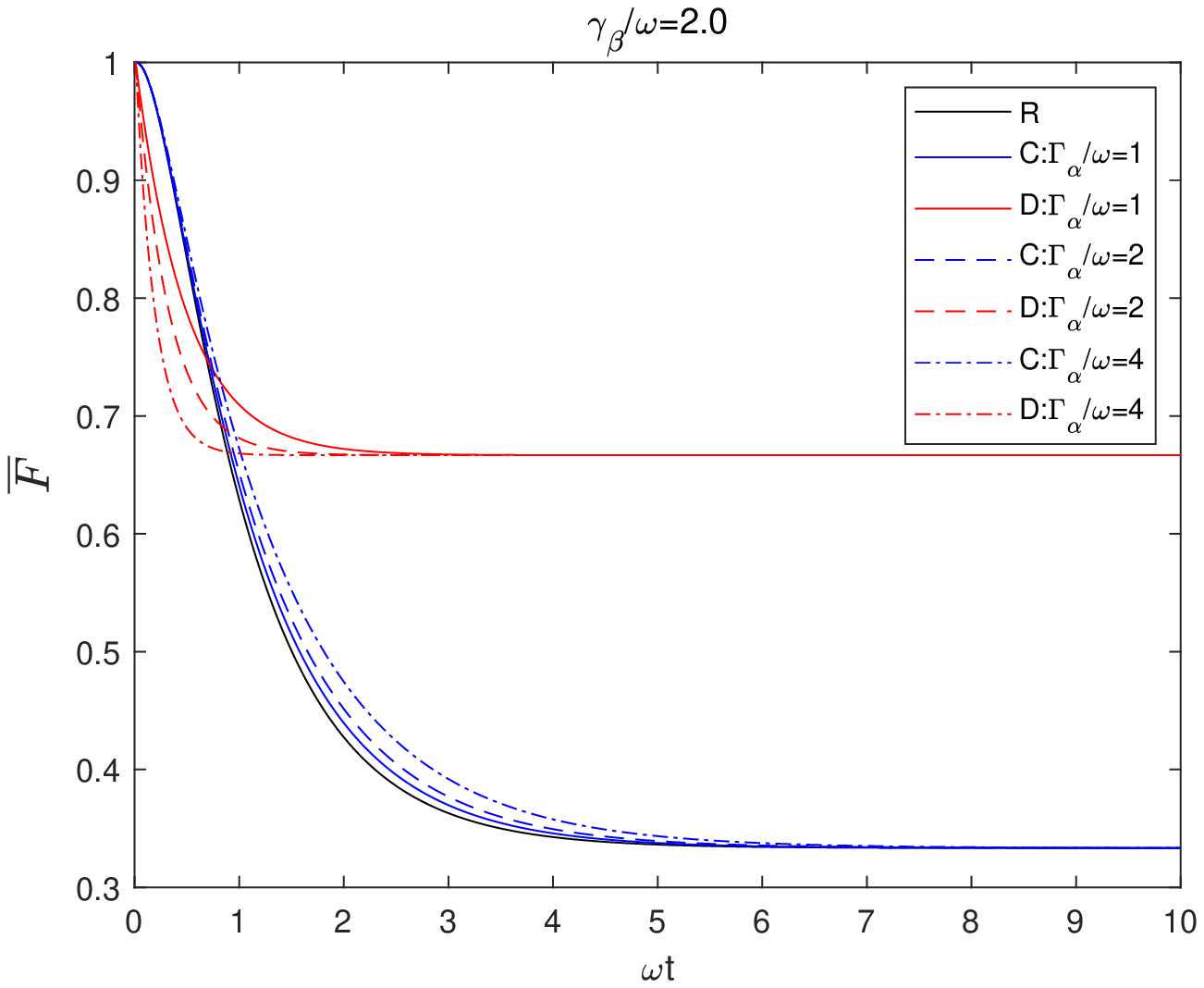}\label{fig3b}fig.3b

\caption{(Color online) Average fidelity $\overline{\mathcal{F}}$ of a two-qubit
state influenced by Markovian dephasing noise in the non-Markovian
relaxation process with various memory capacities: (a)$\gamma_{\beta}/\omega=0.1,$
(b)$\gamma_{\beta}/\omega=2.0$. R, C, and D correspond to the dynamics
under pure non-Markovian relaxation, non-Markovian composite and pure
dephasing processes, respectively. $\Gamma_{\beta}/\omega=1,2,4$
is a fixed parameter representing the dephasing noise coupling strengths,
drawn with solid, dashed and dot-dashed lines. \label{fig3}}
\end{figure}

In order to study how the dephasing noise effects relaxation process
on quantum teleportation, we obtain the average fidelity $\overline{\mathcal{F}}$
for these three types of noise scenarios with different inverse memory
capacity $\gamma_{\beta}/\omega=0.1,2.0$, repectively in Fig.\ref{fig3}(a,b).
The Fig.\ref{fig3}a shows the average fidelity dynamics driven by
these three noise types; non-Markovian pure relaxation noise (black
line R), non-Markovian composite noises (blue lines C) and Markovian
dephasing noise (red lines D). Similar to above studies on capacities
of super-dense coding, in Fig.\ref{fig3}a the average fidelity of
two-qubit system with a strong non-Markovian relaxation processes,
when the memory capacity $\gamma_{\beta}/\omega=0.1$, all three different
colored curves with the same dephasing coupling strength $\Gamma_{\alpha}$
can be divided in to three regions. In a moderate-time scale $0<\omega t<\tau_{1}$,
with the value $3.5<\tau_{1}<5.5$, the system under the mixture noises
have better average fidelity than the systems with pure relaxation
R or pure dephasing D. The second region is $3.5<\omega t<\tau_{2}$,
with $5.0<\tau_{2}<6.5$, that the average fidelity of mixture noises
C are higher than the pure non-Markovian noise R, but smaller than
the pure dephasing noise D. The third is in $6.5<\omega t$, that
mixture noises C smaller than both other two noise types. As for the
rest of the period, mixture noises C smaller than other two noise
types for long-time scale. We can clearly see that among these three
regions: the proposed noise control protocol works well to mitigate
both pure Markovian dephasing noise process and pure non-Markovian
relaxation process at the first region, and works only for pure non-Markovian
relaxation not for dephasing noise. The average fidelity improves
noticeably with the higher dephasing coupling strength $\Gamma_{\alpha}$,
when there is strong dephasing coupling strength $\Gamma_{\alpha}/\omega=4$,
result in way better average fidelity and more delay of the $\tau_{1}$
and $\tau_{2}$. 

When the inverse memory capacity increases to near Markovian $\gamma_{\beta}/\omega=2.0$
in Fig.\ref{fig3}b, the average fidelity $\overline{\mathcal{F}}$
of teleportation for all these noise type scenarios decay significantly
faster than in strong non-Markovian $\gamma_{\beta}/\omega=0.1$.
The average fidelity $\overline{\mathcal{F}}$ dynamics evolves in
the similar pattern as in Fig.\ref{fig3}a, but at much faster rate,
the values of $\tau_{1}$ and $\tau_{2}$ shortens significantly.
Thus, indicate the fewer backflow of information makes differences
between the Fig.\ref{fig3}a and Fig.\ref{fig3}b, as a property of
a non-Markovian process. 

The above works show that, the degeneration of open quantum systems
are inevitable, but by inducing Markovian dephasing noise with the
introducing magnetic field in the non-Markovian relaxation processes,
the two-qubit systems have better average fidelity $\overline{\mathcal{F}}$
and super-dense coding capacities $\chi$, compare to the two-qubit
systems with pure non-Markovian pure relaxation noise processes or
pure dephasing processes in the moderate-time scale. Jing \cite{Jing(2018)}
showed that the average fidelity of single-qubit can be improved by
mixing the dephasing noise with the non-Markovian relaxation noise
processes, and he proposed a control principle: To mutually cancel
two unwanted noisy processes enforced in the unitary evolution of
the quantum system, the characteristic time scales of these two processes
should be separable. In our studies on average fidelity $\overline{\mathcal{F}}$
of teleportation and capacity $\chi$ of super-dense coding in the
two-qubit cases in (Fig.\ref{fig1},\ref{fig2},\ref{fig3}), with
the increase of  non-Markovian memory capacites $1/\gamma_{\beta}$,
more information backflow into quantum systems and strengthen the
correlation between qubits by restoring the entanglement that lost
in the environment. And with the increase the strecngth of magnetic
field leads with high dephaing noise coupling strength $\Gamma_{\alpha}$
in non-Markovian composite noises process, it improves the efficiency
of teleportation and quantum super-dense coding comparably than other
two types of noise scenarios. Since our target is two-qubit quantum
system, these noises are chosen to be enforced on qubits, simultaneously,
and with equal strengths; thus, when measurements are performed, $\overline{\mathcal{F}}$
and $\chi$ are treated equally. This quantum noise control protocol
does not need to perform measurements in the way of Zeno noise control\cite{Sudarshan(1997)},
in fact just by apply a magnetic field, that introduce dephasing noise
and increase its coupling strength, the quantum systems correlation
increases, enhance super-dense coding capacity and fidelity of quantum
teleportation compared to when there is only one types of noise cases. 

\section{conclusion}

In this work, we studied the quantum teleportation and super-dense
coding in the open two-qubit system under the non-Markovian approximation,
and the environmental noises are chosen as three distinct types: Markovian
pure dephasing noise, non-Markovian pure relaxation noise, and mixtures
of these noises. For simplicity, these noises are from two separate
baths. By comparing these three groups of results for average fidelity
$\overline{\mathcal{F}}$ and capacity $\chi$, we found that by introducing
strong magnetic field to induce dephasing noise as a noise control
protocol on the two-qubit non-Markovian relaxation processes, we successfully
enhanced the average fidelity $\overline{\mathcal{F}}$ of teleportation
and capacity $\chi$ of super-dense coding in some time-scale. At
the same time, by studying different strengths of non-Markovian memory
time scales $1/\gamma_{\beta}$, we prove that when there is a strong
non-Makovian process, decoherence delayed significantly and vice versa,
which is proven by others on separate papers for different cases \cite{Chen(2014),Jing(2013),Jing(2015),Jing(2018),L.Di=0000F3si(1998),LuoS.L.(2010),T.Yu(1999),T.Yu(2004),YuTing2004,ZhaoXY(2011)}.
Our findings emphasize that, when two-qubit in an environment of combined
semi-classical and non-Markovian noises, the two noises cancel each
other out and suppressed the non-Markovian open quantum system early
in the process, for that the extra magnetic is needed. The extra magnetic
field helps to reduce the coupling strength $\widetilde{\Gamma}_{\beta}$
and increase the inverse memory capacity $\tilde{\gamma_{\beta}}$
of composite non-Markovian OU noise, the relation between these two
parameters are need to be further study. 
\begin{acknowledgments}
This work is supported by Key Research and Development Plan of Ministry
of Science and Technology, China (No. 2018YFB1601402) and also supported
by the National Natural Science Foundation of China (Grant No.11864042).
\end{acknowledgments}

\end{document}